\documentclass[11pt]{article}
\usepackage{graphicx}
\begin{document}
\bibliographystyle{unsrt}
\def\ra{\rangle}
\def\la{\langle}
\def\aao{\hat{a}}
\def\aaot{\hat{a}^2}
\def\aco{\hat{a}^\dagger}
\def\acot{\hat{a}^{\dagger 2}}
\def\ano{\aco\aao}
\def\bao{\hat{b}}
\def\baot{\hat{b}^2}
\def\bco{\hat{b}^\dagger}
\def\bcot{\hat{b}^{\dagger 2}}
\def\bno{\bco\bao}
\def\beqn{\begin{equation}}
\def\eeqn{\end{equation}}
\def\bear{\begin{eqnarray}}
\def\eear{\end{eqnarray}}
\def\cdott{\cdot\cdot\cdot}
\def\bcen{\begin{center}}
\def\ecen{\end{center}}
\title{Photon-added Coherent States in Parametric 
Down-conversion}
\author{
S. Sivakumar\\Materials Physics Division\\ 
Indira Gandhi Centre for Atomic Research\\ Kalpakkam 603 102 INDIA\\
Email: siva@igcar.gov.in\\
Phone: 91-044-27480500-(Extension)22503}
\maketitle
\begin{abstract}
Photon-added coherent states have been realized in optical parametric 
down-conversion by Zavatta {\em et al} [Science 306 (2004) 660-662].  In this 
report, it is  established  that the states generated in the process are 
{\em ideal} photon-added coherent states.  It is shown that the scheme can 
generate higher order photon-added coherent states.  A comparative study of 
the down-conversion process and atom-cavity interaction in generating the  
photon-added coherent states is presented.   

\end{abstract}
\newpage
\section{Introduction}\label{secI}

Quantum-classical divide continues to be enigmatic. Good experiments are 
necessary to improve our understanding of the issue.  One way is to generate 
and study states that interpolate between the classical and quantum domains.   
Coherent state is considered to be a classical state in the sense that the 
Glauber-Sudarshan function is an admissible phase-space probability density 
distribution, {\em i.e.}, non-negative on the entire phase space.  This feature 
is retained by coherent states of arbitrary amplitude.   Additionally, coherent 
states exhibit minimum fluctuations(uncertainties)  in their amplitudes and 
phases.  States obtained by the action of harmonic oscillator creation operator 
on the coherent states do not admit non-negative phase-space distributions.  Such 
states are classified as nonclassical.  Photon-added coherent states (PACS) 
is defined as $\vert\alpha,m\ra=\aao^{\dagger m}\vert\alpha\ra$ (unnormalized), 
where $\vert\alpha\ra$ is a coherent state of amplitude $\alpha$,  $\aco$ is 
the  creation operator for the field and $m$ is a nonnegative 
integer\cite{gsatara}.  The state $\vert\alpha,m\ra$ is said to be a PACS of 
order $m$ and amplitude $\alpha$.  
Unlike the coherent states $\vert\alpha\ra$, the PACS $\vert\alpha,m\ra$ 
exhibits nonclassical features such as squeezing, sub-Poissonian statistics, etc
\cite{gsatara}.  Hence, the PACS is considered as a state that interpolates 
between the classical and nonclassical states.  While the action of the creation 
operator on
a coherent state leads to a nonclassical state, the action of the annihilation 
operator $\aao$ on a  coherent state does not change the state.  Thus, experimentally 
realizing the action  of the creation and annihilation operators on a  
state is a way of testing the fundamental commutation relation 
$\left[\aao,\aco\right]=I$ \cite{zavatta, parigi,dodonov}.  More over, 
photon addition to any state, not necessarily the  coherent state, of light is being 
viewed as a way of introducing nonclassicality\cite{parigi}.\\

 Generation of the PACS is possible in the     
cavity-atom interaction\cite{gsatara}, conditional measurements of beam-splitter 
output\cite{dakna}, etc.  Single-photon-added coherent 
states (SPACSs) have been  experimentally realized in an all-optical scheme employing a 
nonlinear medium\cite{zavatta}.  It has been  shown that the higher order PACS 
$\vert\alpha,m\ra$   corresponding to $m>1$,  can 
be realized in this scheme\cite{gerry}.  It may be added  that the recent suggestion for 
tailoring the  interaction in optomechanical systems (micro-resonator interacting with 
laser light) by proper detuning can be used to generate PACS in the optomechnical 
domain\cite{optojosa}.  A common feature of the aforementioned systems is that they 
are all interacting, bipartite systems.  During evolution, the two subsystems are 
entangled. This enables to make suitable conditional measurements on one of the 
subsystems so that the other 
subsystem   is prepared in a PACS.  Except for the cavity-atom scheme, the other  
proposals are based on bipartite, coupled oscillators. For instance, in the 
optomechanical scheme, the resonator mode is an oscillator and the laser field 
is  another oscillator. Interaction between the oscillators can be tailored by 
detuning.  In particular,  the following forms of interactions are possible
\cite{LYS}:
\bear
H_{n}&\propto&\aco\bco+\aao\bao,\\
H_{p}&\propto&\aco\bao+\aao\bco.
\eear 
In the optomechanical case, the interactions 
 $H_{n}$ and $H_{p}$ correspond respectively to negative detuning when the resonator 
frequency is smaller than the laser frequency and positive 
detuning wherein the laser 
frequency is smaller than the resonator frequency.   The operators $\aco$ 
and $\bco$  are the creation operators of the quantized 
optical field and the resonator mode respectively.  The corresponding annihilation 
operators are $\aao$ and $\bao$ respectively.  
The Hamiltonian $H_n$ 
describes the opto-mechanical equivalent of the all-optical system  
in the experimental scheme of Zavatta {\it et al}.  Similar interaction 
Hamiltonians arise in the context  of optically coupled  nano-resonators
\cite{zhang} and optical parametric amplifiers\cite{LYS}.  
\\

In this report, a comparative study of two processes, namely, parametric downcoversion and atom-cavity interaction,  that can generate PACS is presented.  
By expressing the time-evolved states in a 
suitable non-orthogonal basis, it is established that the later method generates 
{\em ideal} PACS, a feature that is not present in the atom-cavity scheme.  
Further, the parametric downconversion itself is shown to be capable 
of  generating {\em ideal} $m$-photon-added coherent state (MPACS), 
without requiring higher order processes.  

\section{Cavity-atom scheme}

      In the cavity-atom scheme, a two-level atom interacts with the electromagnetic 
field in a single mode cavity. The two levels of the atom are $\vert g\ra$ and 
$\vert e\ra$ respectively.  Interaction between the two-level atom and  the field 
mode of the cavity  is described by the Jaynes-Cummings Hamiltonian\cite{berman} 
\beqn
\hat{H}_{JC}=\hbar\beta\left[\aco\vert g\ra\la e\vert+\aao\vert e\ra\la g\vert\right].
\eeqn
Here $\aco$ and $\aao$ are respectively the creation and annihilation operators for the 
 qunatized field in the cavity. The transition operators  
$\vert g\ra\la e\vert$ and $\vert e\ra\la g\vert$ are respectively the 
lowering and raising operators for the atom.  The coupling strength between the 
atom and the cavity field is characterized by the coupling constant $\beta$.   
The initial state of the system is $\vert\alpha\ra\vert e\ra$, {\em i.e.}, the state of 
cavity field is the  coherent state  $\vert\alpha\ra$ 
and the  atom  is in the   excited state  $\vert e\ra$.  
For short times, the evolution operator 
$\exp\left[-it\hat{H}_{JC}/\hbar\right]$ can be truncated to first order in $\beta t$.  
In this approximation,  the state of the system at time $t$ is 
\beqn
\vert\psi(t)\ra_{\hbox{app}}\approx\vert\alpha\ra\vert e\ra-i\beta t\aco\vert\alpha\ra\vert g\ra.
\eeqn
The approximate final state  $\vert\psi(t)\ra_{\hbox{app}}$ is an entangled state of the 
cavity field and the atom.  If the atom is detected in its ground state 
$\vert g\ra$, the cavity field is the SPACS $\aco\vert\alpha\ra$ 
(not normalized), whose amplitude is same as that of the initial coherent state $\vert\alpha\ra$.  
The suitability of this approximation depends on the 
interaction duration $t$ and  the coupling constant $\beta$.  To assess the 
nature of the approximation in a better way,   the complete time-evolved state 
is required.  To this end,  the evolution operator is expanded as a power series 
in $\hat{H}_{JC}$.  Using the series expression for the evolution operator ,  
the time-evolved state is expressed as 
\beqn
\vert\psi(t)\ra=\sum_{n=0}^{\infty} \frac{\tau^{2n}}{(2n)!}(\aao\aco)^{n}
\vert\alpha\ra\vert e\ra+\sum_{n=0}^\infty\frac{\tau^{2n+1}}{(2n+1)!}
(\aao\aco)^{n}\aco\vert\alpha\ra\vert g\ra.
\eeqn
where $\tau=-i\beta t$. The states $\vert e\ra$ and $\vert g\ra$ are orthogonal to 
each other.  Therefore, if the atom is in the ground state $\vert g\ra$ on exit from the cavity, the state of the field 
in the cavity is 
\beqn
\vert\alpha,\tau\ra\ra=\sum_{n=0}^\infty\frac{\tau^{2n+1}}{(2n+1)!}(\aao\aco)^{n}
\aco\vert\alpha\ra.
\eeqn
The symbol $\vert\dots\ra\ra$ denotes the state of the cavity field.
Using the identities\cite{gsawolf}, 
\bear
(\aco\aao)^n&=&\sum_{k=1}^n S(n,k)\aao^{\dagger k}\aao^k, \\
\aao^{\dagger k}\aao^k\aco&=&k\aao^{\dagger k}\aao^{k-1}+\aao^{\dagger k+1}\aao^k,
\eear
where $S(n,k)$ are the Stirling numbers of the second kind\cite{grad}, the expression 
for $\vert\alpha,\tau\ra\ra$ is recast as 
\beqn\label{cavity}
\vert\alpha,\tau\ra\ra=\sum_{m=1}^\infty\sum_{n=1}^\infty
\frac{\tau^{2n+1}}{(2n+1)!}B(n,m,\alpha)\vert\alpha,m\ra.
\eeqn 
Here, $$B(n,m,\alpha)=\alpha^{m-1}[mS(n,m)+S(n,m-1)m!\alpha]\sqrt{m!
L_m(-\vert\alpha\vert^2)}$$ 
and $\vert\alpha,m\ra=\aao^{\dagger m}\vert\alpha\ra$, are the MPACS 
(unnormalized)of amplitude $\alpha$.   The function $L_m(-\vert\alpha\vert^2)$ is 
the  Laguerre function of order $m$\cite{grad}.  The above result implies that the cavity field is a 
superposition of various MPACS.  The MPACS of different orders but of same 
amplitude  are 
linearly independent and non-orthogonal.  Hence, the superposition 
coefficients in Eq. \ref{cavity} cannot be interpreted as probability 
amplitudes.

\begin{figure}
\centering
\includegraphics[height=10cm,width=10cm]{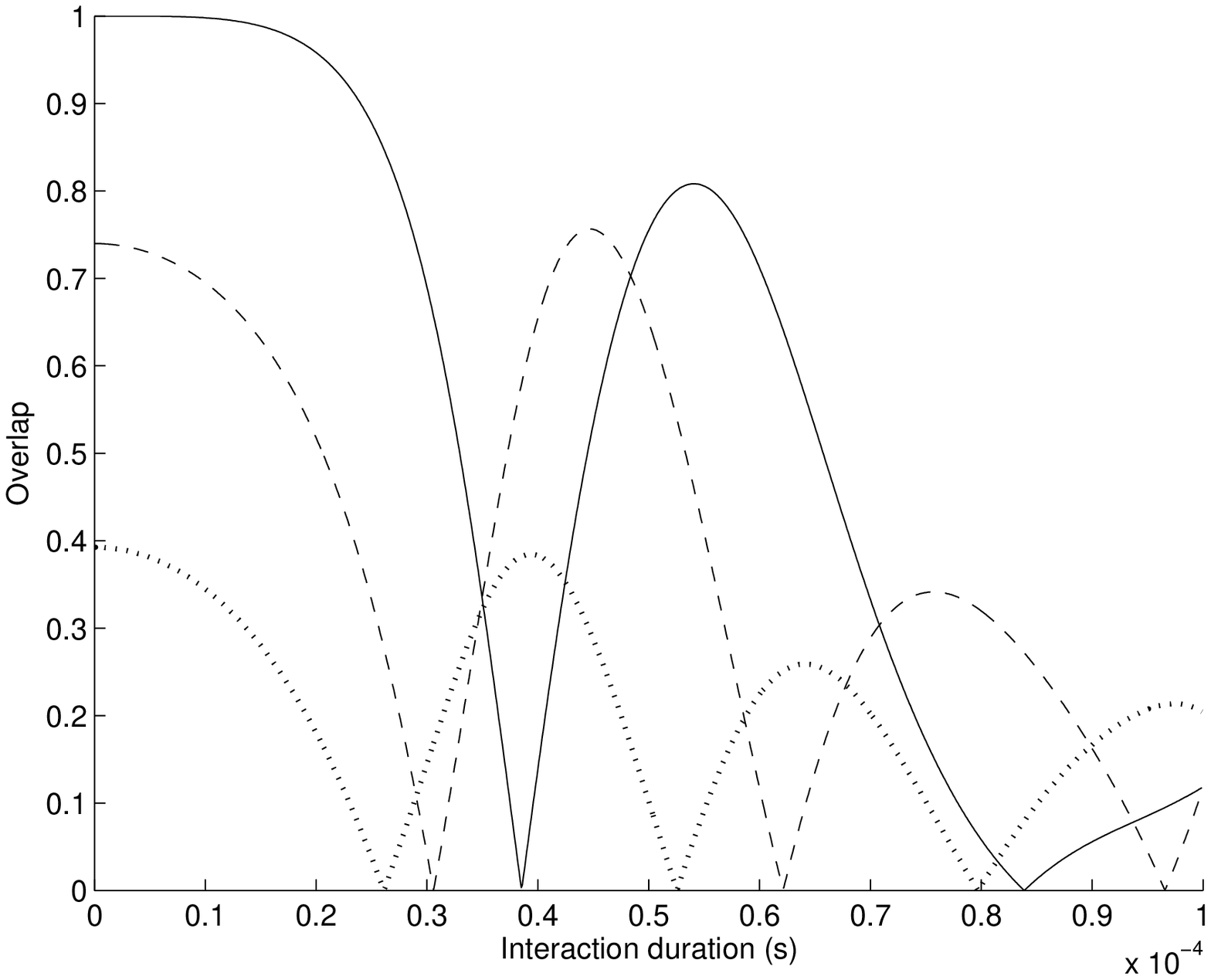}
\caption {Temporal evolution of overlap $\vert\la\alpha,m\vert\alpha,\tau\ra\ra\vert^2$ of the state of the cavity and the MPACS is shown for 
$m$=1 (continuous), 2 (dash) and 3 (dot).  The initial state is a coherent state of amplitude 
$\alpha=0.8$.  }
\label{fig:one}
\end{figure}

On detecting the atom in its ground state,  
the cavity field is expected  to be in  the SPACS $\vert\alpha,1\ra$.    
If it is 
indeed the case, it is not enough that the overlap between the states 
$\vert\alpha,\tau\ra\ra$ and $\vert\alpha,1\ra$ is nearly unity.  It is required 
that $\vert\la\la\alpha,\tau\vert\alpha,m\ra\vert\approx
\vert\la\alpha,1\vert\alpha,m\ra\vert$ for all $m$ to ensure that the states 
generated are indeed the PACS of amplitude $\alpha$.  In Fig. \ref{fig:one}, 
the variation of overlap as the interaction duration increases is shown.          
The coupling constant $\beta$ is chosen to be $2\pi$ MHz.  The respective 
overlap functions of the cavity state $\vert\alpha,\tau\ra\ra$ with 
the MPACS of order $m=1,2$ and 3 are shown.  
The initial coherent state is of amplitude $\alpha=0.8$ and it is expected that the 
scheme generates the SPACS of amplitude $\alpha=0.8$.        
As expected, for short interaction times the overlap between the cavity state
$\vert\alpha,\tau\ra\ra$ and 
the SPACS $\vert\alpha,1\ra$ remains close to unity 
and continues to be substantial $(>0.9)$ even when the interaction duration 
extends to  30$\mu$s.   
However, the overlap of the cavity state with higher order PACS are much 
smaller than the required values.
For instance, the overlap with 
$\vert\alpha,2\ra$ (dashed line in Fig. \ref{fig:one}) falls to 0.3 from 
the short time value of 0.74  if the interaction duration extends to 
30 $\mu$s.  Similarly, overlap with $\vert\alpha,3\ra$ (dotted curve in Fig. 
\ref{fig:one}) decreases rapidly with the increase of interaction time.  
In short, as the interaction time becomes longer, the state of the cavity field  
differs  significantly from the expected SPACS $\vert\alpha,1\ra$.

\section{Coupled oscillators scheme}

In the coupled oscillators scheme, the bipartite system is composed of two 
oscillators which interact. The two oscillator systems could be the two 
modes of the electromagnetic field or the field mode of a microcavity and 
a laser field or  two coupled microresonators,  etc.  In this 
work, the two oscillators are referred as $a$-mode and $b$-mode respectively.  The 
creation and annihilation operators for the $a$-mode are 
$\aco$ and $\aao$ respectively.  The corresponding operators for the $b$-mode are 
$\bco$ and $\bao$ respectively.  The Hamiltonian describing the 
interaction between the two modes in this bipartite system is 
\beqn
\hat{H}_n=\hbar\lambda\left(\aco\bco+\aao\bao\right),
\eeqn
where $\lambda$ is the coupling constant. For short times, the corresponding evolution operator 
$\exp(-i\lambda t\hat{H}_n/\hbar)$ is truncated to $1-i\lambda t\hat{H}_n$.  If the 
initial state of the two modes is $\vert\alpha\ra\vert 0\ra$, then the evolved 
state is a superposition of $\vert\alpha\ra\vert 0\ra$ and 
$\vert\alpha,1\ra\vert 1\ra$.  On detecting the $b$-mode in the one-photon state 
$\vert 1\ra$, the $a$-mode is prepared in the SPACS $\vert\alpha,1\ra$. \\

In order to know how well the generated state approximates the SPACS, it is necessary  
to solve for the dynamics without making any approximation.  This is facilitated by 
the fact that the operators $\aco\bco$, $\aao\bao$ and $\ano +\bno$ provide a realization of 
the generators of the SU(1,1) algebra.  
Consequently,  the evolution operator $\exp(-it\hat{H}_n/\hbar)$ admits the following  
factorization\cite{yamamoto},   
\beqn
\exp(-i\lambda t(\aco\bco+\aao\bao)=\exp(u\aco\bco)\exp[v(\ano+\bno+1)]
\exp(w\aao\bao).
\eeqn
With $\lambda t=r\exp(i\phi)$, the constants $u,v$ and $w$ are 
$\tanh r$,$-\log\cosh r$ and $-\tanh r$ respectively. This factorized form of the  
of the evolution operator is used to obtain the state of the 
bipartite system at time $t$.  If the initial state is $\vert\alpha\ra\vert 0\ra$, 
the state $\vert\chi\ra$ of the coupled oscillators at time $t$ is
\beqn
\vert\chi\ra=\frac{\exp(-\frac{\vert\alpha\vert^2\tanh^2r}{2})}{\cosh r}\sum_{n=0}^\infty\frac{(-i\exp(i\phi)\tanh r)^n}{\sqrt{n!}}
\aao^{\dagger n}\vert\tilde\alpha\ra\vert n\ra.
\eeqn 
In the above expression,  the state $\vert\tilde\alpha\ra=\vert\alpha/\cosh r\ra$ 
 is a coherent state of amplitude $\alpha/\cosh r$.  The state 
$\vert\chi\ra$ is an entangled state of the two modes.  The Fock states of the 
$b$-mode appearing in the expression for $\vert\chi\ra$ are orthogonal to 
each other while the states $\aao^{\dagger n}\vert\tilde\alpha\ra$ of the $a$-mode 
are not.  The orthogonality of the states of the $b$-mode renders it  possible to 
make  
conditional measurements so that the state of the other mode is 
the PACS.  In particular, if  $b$-mode is in the one-photon state $\vert 1\ra$, the 
state of the $a$-mode is the SPACS  
$\vert\tilde\alpha,1\ra=\aco\vert\tilde\alpha\ra$.   More generally, 
if $b$-mode is measured to be the 
number state $\vert m\ra$, then state of the $a$-mode is 
$\aao^{\dagger m}\vert\tilde\alpha\ra$ which is MPACS.  
If photon losses  due to absorption and other dissipative mechanisms are absent or 
negligible and  the interaction duration is sufficiently longer so that 
$r=\vert\lambda\vert t>>1$, then the amplitude $\tilde\alpha$ of the state of the 
$a$-mode becomes nearly zero as $\cosh r$ becomes large.   
In this limit, the state of the $a$-mode is very close to the 
number state $\vert m\ra$.\\

A major difference between the states generated in the cavity-atom 
interaction and those generated in the coupled oscillators is worth mentioning.    
In the later scheme,  the state of the $a$-mode is precisely the SPACS of 
amplitude of  $\alpha/\cosh r$  if the $b$-mode is in its first excited state.  
In general, if the $b$-mode is 
detected to be in the Fock state $\vert m\ra$, the state of the $a$-mode is 
the MPACS $\vert\tilde\alpha,m\ra$.
In the atom-cavity case, the 
state of the cavity  has 
contributions from the PACS of all orders of amplitude $\alpha$.  This superposition 
is never an ideal PACS.\\

The overlap between the SPACS $\vert\tilde\alpha,1\ra$ generated in the process of 
downconversion and the expected SPACS $\vert\tilde\alpha,1\ra$ is  
\beqn
\vert\la\alpha,1\vert\tilde\alpha,1\ra\vert^2=\left[\frac{1+\frac{\vert\alpha\vert^2}
{\cosh^2r}}{1+\vert\alpha\vert^2}
\right]
\exp\left[-\vert\alpha\vert^2(1-\frac{1}{\cosh^2r})\right].
\eeqn
The overlap between  saturates at 
$\exp(-\vert\alpha\vert^2)/(1+\vert\alpha\vert^2)$ as $t\rightarrow\infty$.   
Hence, in this scheme too the overlap of the generated state with SPACS of amplitude 
$\alpha$ falls with interaction duration.  However, as noted previously, the states generated 
are indeed PACS of suitably scaled amplitude.\\

The coefficient in the expression for $\vert\chi\ra$ is the probability amplitude 
for realizing the state $\vert\tilde\alpha,m\ra\vert m\ra$.  Hence, the relavant 
 probability is 
\beqn
P_m=\vert\la\chi\vert\aao^{\dagger m}\vert\tilde\alpha\ra\vert m\ra\vert^2=\frac{\exp(-\vert\alpha\vert^2\tanh^2r)}
{\cosh^2r}{\tanh^{2m}r}L_m\left(-\frac{\vert\alpha\vert^2}{\cosh^2 r}\right).            
\eeqn
Since different Fock states of the $b$-mode are orthogonal to each other, 
the probability $P_m$ is the probability of realizing the MPACS 
$\aao^{\dagger m}\vert\alpha\ra$.  Though this probability decreases with increasing $m$,   
 there is a finite probability of detecting the states corresponding to higher 
values of $m$.  In practical terms, this would mean that more experimental runs 
will be required.   Nevertheless, the method of Zavatta {\it et al } can 
generate {\em ideal} MPACS.\\


\section{Summary}

States generated in the parametric downcoversion process are 
{\em ideal} photon-added coherent states.  The amplitude of the photon-added 
coherent state generated 
in the process is smaller in magnitude compared to the amplitude of the 
 initial seed coherent state.  
If the initial coherent state is of amplitude $\cosh(\lambda\tau)\alpha$, 
 the photon-added coherent state generated is of amplitude $\alpha$. 
This relation fixes the amplitude of the 
initial coherent state in terms of the interaction duration $(\tau)$, coupling constant 
($\lambda$) and the required amplitude for the photon-added coherent state.  
The process is capable of generating ideal $m$-photon-added coherent states, though 
the probability of generation falls with increasing $m$.  In contrast, the 
interaction between a two-level atom and a cavity field in a coherent state 
does not  generate ideal photon-added coherent state of any amplitude.  
In the 
atom-cavity scheme, higher order processes are required to generate higher 
order photon-added coherent states. Typically, $m$-photon processes are 
necessary for 
generating $m$-photon-added coherent states.\\

{\bf Acknowledgement}\\

The author is grateful to  Prof. G. S. Agarwal for useful discussions.  

\newpage

\end{document}